\documentclass[sigconf,nonacm]{acmart}

\AtBeginDocument{%
  }

\setcopyright{none}
\acmConference[CCS '24]{This paper was accepted at the 2024 ACM SIGSAC Conference on
Computer and Communications Security}{October 14--18, 2024}{Salt Lake City,
UT, USA}
\acmDOI{10.1145/3658644.3691391}

\usepackage{todonotes}

\newcommand{\mypar}[1]{\smallskip\noindent\textbf{#1.}}

\usepackage{hyperref}
\usepackage{soul}
\usepackage{svg}

\usepackage[english]{babel}
\title{Poster: libdebug, Build Your Own Debugger for a Better \\ (Hello) World}
\thanks{This paper has been accepted at the 2024 ACM SIGSAC Conference on Computer and Communications Security (CCS '24), October 14--18, 2024, Salt Lake City, UT, USA}

\begin{document}

\author{Gabriele Digregorio}
\orcid{0009-0003-1854-759X}
\email{io_no@libdebug.org}
\email{gabriele.digregorio@polimi.it}
\affiliation{%
  \institution{Politecnico di Milano}
  \city{Milano}
  \country{Italy}
}
\author{Roberto Alessandro Bertolini}
\orcid{0009-0002-2314-6056}
\email{mrindeciso@libdebug.org}
\email{robertoalessandro.bertolini@mail.polimi.it}
\affiliation{%
  \institution{Politecnico di Milano}
  \city{Milano}
  \country{Italy}
}
\author{Francesco Panebianco}
\orcid{0009-0007-1510-2594}
\email{frank01001@libdebug.org}
\email{francesco.panebianco@polimi.it}
\affiliation{%
  \institution{Politecnico di Milano}
  \city{Milano}
  \country{Italy}
}
\author{Mario Polino}
\orcid{0000-0002-0925-2306}
\email{jinblack@libdebug.org}
\affiliation{%
  \city{Milano}
  \country{Italy}
}
\renewcommand{\shortauthors}{Gabriele Digregorio, Roberto Alessandro Bertolini, Francesco Panebianco, \& Mario Polino}

\renewcommand{\shorttitle}{Poster: libdebug, Build Your Own Debugger for a Better (Hello) World}

\begin{abstract}
Automated debugging, long pursued in a variety of fields from software engineering to cybersecurity, requires a framework that offers the building blocks for a programmable debugging workflow. However, existing debuggers are primarily tailored for human interaction, and those designed for programmatic debugging focus on kernel space, resulting in limited functionality in userland.
To fill this gap, we introduce libdebug, a Python library for programmatic debugging of userland binary executables. libdebug offers a user-friendly API that enables developers to build custom debugging tools for various applications, including software engineering, reverse engineering, and software security. It is released as an open-source project, along with comprehensive documentation to encourage use and collaboration across the community. We demonstrate the versatility and performance of libdebug through case studies and benchmarks, all of which are publicly available. We find that the median latency of syscall and breakpoint handling in libdebug is 3 to 4 times lower compared to that of GDB. 

\end{abstract}




\maketitle

\section{Introduction}

Debuggers have long been key tools for software engineers, serving as essential instruments for identifying and resolving logical flaws within code. Among the most prominent debugging tools for Linux is the GNU Debugger (GDB)~\cite{stallman_debugging_1988}, known for its set of human-oriented commands designed for interactive debugging. Through step-by-step manual inspection, software developers can spot unexpected behavior and bugs.
Over time, the need for automation in the debugging process has been increasingly recognized, promoting the formalization and systematization of debugging as a hypothesis verification process~\cite{zeller_why_2009, zeller_debugging_2024}. While these methodologies were originally developed within the context of software engineering, the application of debuggers has since extended beyond traditional software development into the cybersecurity context. Security professionals now leverage debuggers to diagnose memory corruption vulnerabilities and assess potential exploitation vectors that could escalate into remote code execution on target systems. Furthermore, debuggers enable dynamic analysis during reverse engineering, allowing for the dissection of intricate procedures by inspecting the process state in real-time. This capability is essential for tracing the behavior of malicious software and identifying vulnerabilities in binary applications, particularly in scenarios where the source code is unavailable.

A crucial step in automating debugging is creating a programmable framework equipped with essential building blocks to construct tailored debugging workflows. 
An illustrative example of such a framework is drgn~\cite{sandoval_drgn_2024}, a programmable debugger specifically targeted for the Linux Kernel. A core aspect of its design is intuitiveness in scripting, which is a shortcoming for existing solutions like GDB.
However, security specialists frequently need to script and automate debugging within user space to replicate specific execution flows or to develop customized wrappers suited to unique debugging tasks. Although drgn aims to support userland debugging, it was not initially intended for this purpose. Notably, at the time of writing, it can access a userspace process and read its memory, but it lacks fundamental features for effective debugging, including pausing threads and setting breakpoints~\cite{sandoval_live_2023}. 

To address these needs, we introduce \textbf{libdebug}~\cite{libdebug_2024}, an open-source Python library for programmatic debugging. libdebug provides a comprehensive set of building blocks designed to facilitate the development of debugging tools for different purposes, including reverse engineering and exploitation. During the development of libdebug, we prioritized efficiency in the implementation of each building block. This aspect is not secondary, as it directly impacts the execution time of the debuggee and, therefore, the applicability of the designed debugging workflows in real-world scenarios.

This paper presents the rationale behind libdebug, demonstrating its utility across various domains via detailed case studies, with an emphasis on security perspectives. 
To show how libdebug's building blocks can come together, we release the source code of these case studies.
Additionally, we benchmark relevant event handling in both libdebug and GDB Python scripts, specifically breakpoint and syscall handling, and find that using libdebug results in a speedup of 3 to 4 times in the median case.

Our contributions are summarized as follows:

\begin{itemize}
\item We introduce libdebug: a versatile Python library for programmatic debugging of userland binary executables.
\item We demonstrate the capability and potential of libdebug through various use cases, ranging from software engineering to cybersecurity. Additionally, we compare its event handling latency with that of GDB.
\item We release libdebug as an open-source project\footnote{\url{https://github.com/libdebug/libdebug}}, accompanied by comprehensive documentation\footnote{\url{https://docs.libdebug.org}}.
\end{itemize}
\section{Your Own Debugger}

Debugging a userland binary executable involves kernel-mediated control over the state of the target process, such as its memory and register contents.
For example, in the case of Unix-like kernels (e.g., Linux, Darwin), the \textit{ptrace}~\cite{noauthor_ptrace_1999} system call is used to interact with running processes, through a series of architecture-specific commands.
The role of a debugger is to add a layer of abstraction that manages the interaction with these unintuitive native debug APIs, allowing users to focus on the debugging of applications.

Widespread debuggers such as GDB are designed primarily for human interaction. However, many tasks require the debugging flow 
to be repeatable and programmable. Moreover, traditional debuggers typically only focus on debugging binaries during development and testing phases. In cybersecurity, a debugger is a vital tool used in reverse engineering and exploitation. During these tasks, it is crucial to consistently and reliably know the execution state of a process. For example, programmatically inspecting how different inputs affect the execution flow or the content of registers and memory is highly beneficial. 

libdebug is a Python library that provides users with all the necessary building blocks to easily program custom debuggers that can be tailored to unique scenarios and applications, including those related to cybersecurity. Python's consistent popularity, versatility, and widespread adoption among tech experts~\cite{stephen_cass_top_2023} make it an ideal choice for this task.
libdebug implements direct interaction with the native debug APIs exposed by the kernel in a way that is transparent to the user, providing a consistent high-level API across different architectures. libdebug does not rely on any assumptions about the compilation options or the structure of the binary being analyzed, nor does it require for any debug information to be embedded in the executable.
The functionalities exposed by libdebug include operations on registers, memory, breakpoints, watchpoints, syscalls, and signals, as well as support for multithreading. Additionally, it exposes interaction with the target's standard input, output, and error streams.

While it is mostly written in Python, libdebug leverages C code bindings~\cite{python_software_foundation_extending_2024} to improve interaction with the operating system and performance. It is designed with modularity in mind, facilitating the easy expansion of its functionalities and support. At the time of writing, libdebug already supports AMD64 and AArch64. Furthermore, it is not constrained by the ptrace paradigm and different debugging interfaces can be seamlessly integrated. For example, support can be added for the gdbstub exposed by QEMU~\cite{bellard_qemu_2005} or the native debug APIs of other operating systems, such as Windows~\cite{microsoft_corporation_debugging_2023}.
\section{Use Cases}
In this section, we present three examples of how libdebug can be used to create a debugger tailored to a precise task. The first focuses on demonstrating how a tool made with libdebug can assist in the reverse engineering and debugging of operations performed by code interpreters. This includes interpreted programming languages, virtual machines, and Just-In-Time recompilation.
The second example illustrates how libdebug can ease the development of tools for automatic vulnerability detection and exploitation.
The last example highlights the benefits of integrating libdebug in unit testing and coverage analysis toolchains.
The same basic blocks used in these examples can be reshaped and combined to apply libdebug to any other task and domain. We publicly release all example code~\footnote{\url{https://github.com/libdebug/libdebug/tree/0.5.4/examples}}.

\mypar{Interpreted Bytecode Debugging}
To reverse engineer or troubleshoot non-native bytecode, knowing the state of the bytecode interpreter can be particularly useful. This knowledge becomes crucial when no reliable debugging mechanisms exist, or when existing methods fail to provide the necessary control over the interpreter's state. To address this, the proposed solution leverages libdebug’s building blocks to craft a debugger tailored for the interpreter.

In the example, libdebug is used for debugging the CPython interpreter binary and the libpython shared library loaded by the interpreter. At runtime, the script dumps each fetched Python opcode and modifies the interpreter's state to change the execution flow — for instance, changing the type of a \textit{binary operation} from sum to subtraction.
This example showcases a high level of automation, including the identification of optimal locations within the shared library to install breakpoints. On breakpoint hit, the libdebug script specifies the operation to perform according to the task's objectives.


\mypar{Vulnerability Detection and Exploitation}
Fuzzing is an automated technique to identify inputs that cause inconsistent or failure states. This method has received significant attention from both software developers and security specialists~\cite{zhu_fuzzing_2022}. However, once fuzzing yields a candidate input, the task of understanding the bug and assessing its potential as a vulnerability falls to the user.
In our example, we demonstrate a use case employing libdebug to develop a simple program for detecting buffer overflows. This tool catches SIGSEGV signals during the execution of a Linux binary. Upon identifying a problematic input, the script performs a detailed post-mortem analysis, examining registers and memory locations impacted by the input. 
Then, it modifies the input to determine the setup that can overwrite the base pointer or even the instruction pointer, leading to a successful stack-pivoting or control flow hijacking. To facilitate remediation, a stack trace is provided, for precise identification of the vulnerable execution flow.



\mypar{Automated Testing and Coverage}
Modern software engineering practices emphasize the need for frequent and extensive testing to quickly detect and correct errors.
Code coverage analysis ensures all edge cases are tested by tracing and analyzing each execution path.
A common coverage metric is the \textit{branch coverage}, which measures the percentage of conditional branches that are taken or not during various executions.
There are two main ways of performing coverage analysis during testing.
Some tools, like GNU gcov~\cite{noauthor_gcov_1996}, employ compile-time binary instrumentation to embed additional code that records any branch taken. Other tools, like kcov~\cite{kagstrom_kcov_2010} and bcov~\cite{neumann_bcov_2007}, perform dynamic analysis to identify any branching path and calculate the corresponding coverage.

Our example shows how libdebug can be used to develop a tool that efficiently calculates the \textit{branch coverage} after each test run. The tool requires minimal development effort. Using breakpoints, it performs dynamic analysis of the code paths taken at runtime.
Moreover, the use case simplifies and enriches the testing process by simulating and validating rare failures, such as those during memory allocation or file access.
These faults can be deterministically injected while running a test case, ensuring their coverage.

\section{Benchmarks}

In the following section, we present two benchmarks that compare the execution times of equivalent tasks using libdebug and GDB. The first benchmark assesses the overhead associated with breakpoint handling, while the second evaluates the overhead in syscall handling.
These two events were chosen for their relevance to many use cases, including those presented in this work, which require efficient programmatic debugging. 

It is important to note that GDB is not primarily designed to be programmatic. Despite this, it offers a Python API for the creation of custom commands that directly interact with the debugger~\cite{stallman_debugging_1988}. Due to their nature, care must be taken while designing custom commands to avoid disrupting the normal functioning of GDB.
Moreover, this approach inevitably incurs additional overhead in printing logs and interpreting commands. Thus, we developed two GDB commands that run the benchmark scripts while disabling as many logs and potential overhead sources as possible.

\begin{figure}[t]
    \centering
    \includegraphics[width=\linewidth]{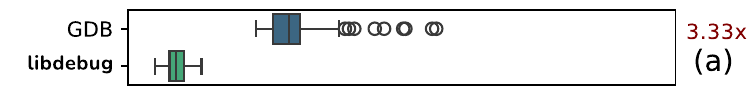}
    \includegraphics[width=\linewidth]{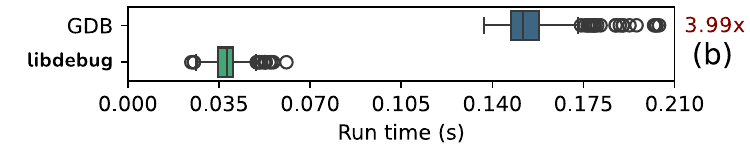}
    \caption{Distributions of run times for handling 1000 breakpoints (a) or 1000 syscalls (b) with libdebug and the GDB Python API. The multiplier on the right is GDB's overhead.}
    \label{fig:benchmarks}
\end{figure}

Each benchmark script handles 1000 events. We ran each benchmark 1000 times on the same machine, equipped with an Intel Core Ultra 7 155H and 64GB of RAM. The kernel version was 6.9.9-arch1-1, while libdebug and GDB were at versions 0.5.4 and 15.1, respectively. All scripts used for the benchmarks are public for reproducibility~\footnote{\url{https://github.com/libdebug/libdebug/tree/0.5.4/test/benchmarks}}. The benchmark results are shown in Figure~\ref{fig:benchmarks}. We can see that in both tests, the run time shown by libdebug is significantly lower than that of GDB, with libdebug being 3 to 4 times faster in the median case.

\section{Conclusions}
Debuggers are essential tools for software engineers and security analysts, but existing solutions often lack the programmability required for advanced automation. 
To address this gap, we developed libdebug, a programmable debugger for userland binary executables. It has been released as open-source software and comes with comprehensive documentation.
libdebug is designed to abstract architecture-specific details, providing unified APIs that can be used across different platforms.
Currently, libdebug supports AMD64 and AArch64 Linux systems, but its modular design enables easy adaptation to other architectures and operating systems in future versions.
This paper demonstrated libdebug's capabilities through various use cases and benchmarks against GDB, with all test source code made publicly available. We find that libdebug is 3 to 4 times faster than GDB at handling some common debugging events.


\bibliographystyle{ACM-Reference-Format}
\bibliography{libdebug_poster_refs}

\end{document}